\documentclass[aps,preprint]{revtex4}%
\usepackage{amsfonts}
\usepackage{amsmath}
\usepackage{amssymb}
\usepackage{graphicx}%
\begin{document}
\preprint{CTP-SCU/2015002}
\title{Minimal Length Effects on Schwinger Mechanism}
\author{Benrong Mu$^{a,b}$}
\email{mubenrong@uestc.edu.cn}
\author{Peng Wang$^{b}$}
\email{pengw@scu.edu.cn}
\author{Haitang Yang$^{b}$}
\email{hyanga@scu.edu.cn}
\affiliation{$^{a}$School of Physical Electronics, University of Electronic Science and
Technology of China ,Chengdu, 610054, China}
\affiliation{$^{b}$Center for Theoretical Physics, College of Physical Science and
Technology, Sichuan University, Chengdu, 610064, PR China}

\begin{abstract}
In this paper, we investigate effects of the minimal length on the Schwinger
mechanism using the quantum field theory (QFT) incorporating the minimal
length. We first study the Schwinger mechanism for scalar fields in both usual
QFT and the deformed QFT. The same calculations are then performed in the case
of Dirac particles. Finally, we discuss how our results imply for the
corrections to the Unruh temperature and the Hawking temperature due to the
minimal length.

\end{abstract}
\maketitle
\tableofcontents

%\affiliation{Center for Theoretical Physics, College of Physical Science and Technology,
%Sichuan University, Chengdu, 610064, PR China}

%\affiliation{Center for Theoretical Physics, College of Physical Science and Technology,
%Sichuan University, Chengdu, 610064, PR China}

\section{Introduction}

Using the proper-time method, Schwinger\cite{IN-Schwinger:1951nm} calculated
the effective action of a charged particle in an external electromagnetic
field. He found that the action has an imaginary part for a uniform electric
field, which leads to the vacuum decay through pair production. Due to its
purely non-perturbative nature, this quantum field theoretical prediction is
of fundamental importance. The Schwinger mechanism sheds lights on topics as
diverse as the string breaking rate in
QCD\cite{IN-Casher:1978wy,IN-Neuberger:1979tb} and on black hole
physics\cite{IN-Brout:1993be}.

On the other hand, various theories of quantum gravity, such as string theory,
loop quantum gravity and quantum geometry, predict the existence of a minimal
length\cite{IN-Townsend:1977xw,IN-Amati:1988tn,IN-Konishi:1989wk}. The
generalized uncertainty principle(GUP)\cite{IN-Kempf:1994su} is a simply way
to realize this minimal length. An effective model of the GUP in one
dimensional quantum mechanics is given
by\cite{IN-Hossenfelder:2003jz,IN-Hassan:2002qk}
\begin{equation}
k(p)=\frac{1}{\sqrt{\beta}}\tanh\left(  \sqrt{\beta}p\right)  , \label{tanh1}%
\end{equation}%
\begin{equation}
\omega(E)=\frac{1}{\sqrt{\beta}}\tanh\left(  \sqrt{\beta}E\right)  ,
\label{tanh2}%
\end{equation}
where the generators of the translations in space and time are the wave vector
$k$ and the frequency $\omega$, $\beta=\frac{\beta_{0}}{m_{p}^{2}}$, $m_{p}$
is the Planck mass and $\beta_{0}$ is a dimensionless parameter marking
quantum gravity effects. We set $c=\hbar=G=1$ in the paper. The quantization
in position representation $\hat{x}=x$ leads to
\begin{equation}
k=-i\partial_{x},\text{ }\omega=i\partial_{t}.
\end{equation}
Therefore, the low energy limit $p\ll m_{p}$ including order of $\frac{p^{3}%
}{M_{f}^{3}}$ gives
\begin{align}
p  &  =-i\partial_{x}\left(  1-\frac{\beta}{3}\partial_{x}^{2}\right)
,\label{eq:momentum}\\
E  &  =i\partial_{t}\left(  1-\frac{\beta}{3}\partial_{t}^{2}\right)  .
\label{eq:energy}%
\end{align}
From eqn. $\left(  \ref{tanh1}\right)  $, it is noted that although one can
increase $p$ arbitrarily, $k$ has an upper bound which is $\frac{1}%
{\sqrt{\beta}}$. The upper bound on $k$ implies that that particles could not
possess arbitrarily small Compton wavelengths $\lambda=2\pi/k$ and that there
exists a minimal length $\sim\sqrt{\beta}$.

In this paper, we investigate scalars and fermions pair production from a
static classical electric field using the deformed QFT which incorporates the
minimal length via eqn. $\left(  \ref{eq:momentum}\right)  $ and eqn. $\left(
\ref{eq:energy}\right)  $. The organization of this paper is as follows. In
section \ref{Sec:SPP}, based on the usual and the minimal length modified
field theoretic considerations, Schwinger's mechanism is derived in the case
of spinless particles. We then show how the same calculations can be performed
in the case of Dirac particles in section \ref{Sec:FPP}. Section \ref{Sec:Con}
is devoted to our discussions and conclusions.

\section{Scalar Pair Production}

\label{Sec:SPP}

In this section, we first derive the formula for the scalar pair production
rate in the framework of QFT. We then use the formula to calculate pair
creation rate in usual QFT\ and the minimal length modified QFT. As\ shown in
ref. \cite{SPP-Holstein:1999ta}, the scalar field theoretic vacuum to vacuum
amplitude can be written as%
\begin{equation}
\left\langle vac\rightarrow vac\right\rangle \propto\int\left[  d\phi\right]
\exp\left(  i\int dx^{4}\mathcal{L}\right)  ,
\end{equation}
where $\mathcal{L}$ is the Lagrange density and $\phi$ is the corresponding
scalar field. Assume $\mathcal{L}$ is given by $\mathcal{L=}\phi
^{+}\mathcal{O}_{s}\phi$ where $\mathcal{O}_{s}$ is some differential
operator. Defining eigenfunctions $\phi_{n}$ with eigenvalues $\lambda_{n},$%
\begin{equation}
\mathcal{O}_{s}\phi_{n}=\lambda_{n}\phi_{n},
\end{equation}
we expand%
\begin{equation}
\phi=\sum\limits_{n}a_{n}\phi_{n}.
\end{equation}
Using orthogonality of $\phi_{n}$, we find%
\begin{align}
e^{-\xi_{s}} &  \equiv\left\langle vac\rightarrow vac\right\rangle \propto\int
da_{n}da_{n}^{\ast}\exp\left(  i\int dx^{4}\lambda_{n}\left\vert
a_{n}\right\vert ^{2}\right)  \nonumber\\
&  \propto\prod\limits_{n}\frac{1}{\lambda_{n}}=\frac{1}{\det\mathcal{O}_{s}%
}=\exp\left(  -\text{Tr}\left(  \ln\mathcal{O}_{s}\right)  \right)
=\exp\left(  -\sum\limits_{n}\ln\lambda_{n}\right)  .\label{eq:xi}%
\end{align}
Using the integral
\begin{equation}
\ln a=-\int_{0}^{+\infty}ds\frac{e^{-as}}{s}+\text{const.},
\end{equation}
one has%
\begin{equation}
\xi_{s}=\xi_{s}^{E}+C,\label{eq:xiC}%
\end{equation}
where $C$ is a constant and we define%
\begin{equation}
\xi_{s}^{E}\equiv-\sum\limits_{n}\int_{0}^{+\infty}\frac{ds}{s}e^{-\lambda
_{n}s}.\label{eq:EtaScalar}%
\end{equation}
The imaginary part of $\xi_{s}$ is always infinite due to vacuum energy shift.
Here, we are only interested in the real part of $\xi^{s}$ which gives the
vacuum decay. As shown later in this section, the electric field $E$ only
appears in $\lambda_{n}$\ and $C$ is independent of $E$. To obtain $C$ we
consider the case without the electric field. When there is no electric field,
the vacuum is stable and no pairs are produced. In this case, the vacuum to
vacuum amplitude is%
\[
e^{i\alpha}=e^{-\xi_{s}}=\exp\left(  -\xi_{s}^{0}+C\right)
\]
where $\alpha$ is a phase irrelevant to the vacuum decay and $\xi_{s}^{0}$ is
$\xi_{s}^{E}$ with $E=0$. Thus, one has
\begin{equation}
C=\xi_{s}^{0}+i\alpha.\label{eq:C}%
\end{equation}
Plugging eqn. $\left(  \ref{eq:C}\right)  $ into eqn. $\left(  \ref{eq:xiC}%
\right)  $, we find that the real part of $\xi_{s}$ is%
\begin{equation}
\gamma_{s}\equiv\operatorname{Re}\xi_{s}=\operatorname{Re}\left(  \xi_{s}%
^{E}-\xi_{s}^{0}\right)  .
\end{equation}
Squaring $\gamma_{s}$, we observe that the total pair production rate per unit
volume is simply%
\begin{equation}
\frac{\text{prob}_{\text{pair}}}{VT}=\frac{1}{VT}\left(  1-e^{-2\gamma_{s}%
}\right)  \approx\frac{2\gamma_{s}}{VT}.\label{eq:ProductionRate}%
\end{equation}

\subsection{Usual Scalar Field}

For the case of a scalar field $\phi$ of mass $m$ and charge $e$ in the
presence of an external electromagnetic interaction described by vector
potential $A_{\mu}$, the Lagrangian is given by%
\begin{equation}
\mathcal{L}_{s}=\phi^{+}\mathcal{O}_{s}^{\left(  0\right)  }\phi,
\end{equation}
where $\mathcal{O}_{s}^{\left(  0\right)  }\mathcal{=}\left(  \partial
+ieA\right)  ^{2}+m^{2}$. If there is a uniform electric field$\ \mathbf{E}%
=E\mathbf{e}_{z},$ we can utilize the gauge $\mathbf{A}=-Et\mathbf{e}_{z}.$
The operator $\mathcal{O}_{s}^{\left(  0\right)  }$ then becomes%
\begin{equation}
\mathcal{O}_{s}^{\left(  0\right)  }\mathcal{=\partial}_{t}^{2}-\partial
_{\perp}^{2}-\left(  \partial_{z}+ieEt\right)  ^{2}+m^{2},
\end{equation}
where $\partial_{\perp}^{2}=\partial_{x}^{2}+\partial_{y}^{2}$. Assume that
eigenfunction of $\mathcal{O}_{s}^{\left(  0\right)  }$ takes the form as
$\psi=\exp\left(  i\mathbf{k}_{\perp}\cdot\mathbf{r}_{\perp}+ik_{z}z\right)
\sigma\left(  t\right)  $ where $\mathbf{r}_{\perp}=x\mathbf{e}_{x}%
+y\mathbf{e}_{y}$ and $\sigma\left(  t\right)  $ satisfies%
\begin{equation}
\left(  \left(  \frac{d}{dt}\right)  ^{2}+k_{\perp}^{2}+\left(  k_{z}%
+eEt\right)  ^{2}+m^{2}\right)  \sigma\left(  t\right)  =\lambda\sigma\left(
t\right)  .\label{EOMscalarno}%
\end{equation}
Performing the Wick Rotation $t\rightarrow-i\tau$ and $E\rightarrow-i\tilde
{E}$, we have%
\begin{equation}
\left(  -\left(  \frac{d}{d\tau}\right)  ^{2}+e^{2}\tilde{E}^{2}\left(
\tau-\frac{k_{z}}{e\tilde{E}}\right)  ^{2}\right)  \sigma\left(  t\right)
=\left(  \lambda-k_{\perp}^{2}-m^{2}\right)  \sigma\left(  t\right)
.\label{eq:Sigma}%
\end{equation}
Obviously, eqn. $\left(  \ref{eq:Sigma}\right)  $ describes a one-dimensional
harmonic oscillator with its well centered at $\frac{k_{z}}{e\tilde{E}}$ and a
resonant frequency $2e\tilde{E}$.\ One can express $\frac{d}{d\tau}$ and
$\tau$ in terms of ladder operators, $a$ and $a^{+}$, as%
\begin{align}
\tau-\frac{k_{z}}{e\tilde{E}} &  =\frac{1}{\sqrt{2e\tilde{E}}}\left(
a^{+}+a\right)  ,\label{eq:Tao}\\
\frac{d}{d\tau} &  =-\sqrt{\frac{e\tilde{E}}{2}}\left(  a^{+}-a\right)
.\label{eq:DTao}%
\end{align}
Thus, the energy levels are quantized as%
\begin{equation}
\lambda_{n,k_{\perp}}^{s\left(  0\right)  }=k_{\perp}^{2}+m^{2}+\left(
2n+1\right)  e\tilde{E}\label{eq: EigenScalar}%
\end{equation}
where $n=0,1,2\cdots$. Note that $k_{x}$ and $k_{y}$ range over all values
from $-\infty$ to $\infty$, but $k_{z}$ is constrained to be in the range
$0<k_{z}<e\tilde{E}iT$ in order that the entire range of time is included as
$k_{z}$ is varied, where $T$ is a total interaction time. Thus, the
corresponding degeneracy is $\frac{eETVdk_{\perp}}{\left(  2\pi\right)  ^{3}}$
where $V$ is the volume. After switching back to $t$ and $E$, eqn. $\left(
\ref{eq:EtaScalar}\right)  $ and eqn. $\left(  \ref{eq: EigenScalar}\right)  $
yields
\begin{equation}
\xi_{s}^{E}=-\frac{eEVT}{\left(  2\pi\right)  ^{3}}\int dk_{\perp}\int
_{0}^{+\infty}\frac{ds}{s}\exp\left[  -\left(  k_{\perp}^{2}+m^{2}\right)
s\right]  \sum\limits_{n=0}^{\infty}\exp\left[  -i\left(  2n+1\right)
eEs\right]  .\label{eq:ESscalar}%
\end{equation}
With the help of Dirac comb
\begin{equation}
\pi\sum_{k=-\infty}^{\infty}\delta\left(  t-k\pi\right)  =\sum_{n=-\infty
}^{\infty}\exp\left(  -i2nt\right)  ,
\end{equation}
one easily gets
\begin{equation}
\operatorname{Re}\xi_{s}^{E}=-\frac{eEVT}{2\left(  2\pi\right)  ^{3}}\int
dk_{\perp}\int_{0}^{+\infty}\frac{ds}{s}\exp\left[  -\left(  k_{\perp}%
^{2}+m^{2}\right)  s\right]  \exp\left(  -ieEs\right)  \pi\sum_{k=-\infty
}^{\infty}\delta\left(  eEs-k\pi\right)  .
\end{equation}
When $E$ approaches zero, we have for $\operatorname{Re}\xi_{s}^{0}$
\begin{equation}
\operatorname{Re}\xi_{s}^{0}=-\frac{VT}{16\pi^{2}}\int dk_{\perp}\int
_{0}^{+\infty}\frac{ds}{s}\exp\left[  -\left(  k_{\perp}^{2}+m^{2}\right)
s\right]  \delta\left(  s\right)  .
\end{equation}
Thus, we find%
\begin{align}
\gamma_{s}^{\left(  0\right)  } &  =\operatorname{Re}\left(  \xi_{s}^{E}%
-\xi_{s}^{0}\right)  \nonumber\\
&  =-\frac{VT}{16\pi^{2}}\int dk_{\perp}\int_{0}^{+\infty}\frac{ds}{s}%
\exp\left[  -\left(  k_{\perp}^{2}+m^{2}\right)  s\right]  \exp\left(
-ieEs\right)  \left[  \sum_{k=-\infty}^{\infty}\delta\left(  s-\frac{k\pi}%
{eE}\right)  -\delta\left(  s\right)  \right]  \nonumber\\
&  =\frac{e^{2}E^{2}VT}{16\pi^{3}}\sum_{k=1}^{\infty}\frac{\left(  -1\right)
^{k+1}}{k^{2}}\exp\left(  -m^{2}\frac{k\pi}{eE}\right)  .
\end{align}

\subsection{Minimal Length Modified Scalar Field}

In the presence of an external electromagnetic potential $A_{\mu}$, eqn.
$\left(  \ref{eq:momentum}\right)  $ and eqn. $\left(  \ref{eq:energy}\right)
$ can be generalized to%
\begin{align}
p_{i} &  =-iD_{i}\left(  1-\frac{\beta}{3}D_{i}^{2}\right)
,\label{eq:MomentumGUP}\\
E &  =iD_{t}\left(  1-\frac{\beta}{3}D_{t}^{2}\right)  ,\label{eq:EnergyGUP}%
\end{align}
where $D_{\mu}=\partial_{\mu}+ieA_{\mu}$. For a scalar field $\phi$ of mass
$m$ and charge $e$ in the external electromagnetic potential $A_{\mu}$, the
Lagrangian incorporating eqn. $\left(  \ref{eq:MomentumGUP}\right)  $ and eqn.
$\left(  \ref{eq:EnergyGUP}\right)  $ can be written as%
\begin{equation}
\mathcal{L}_{s}=\eta^{\mu\nu}\left(  p_{\mu}\phi\right)  ^{+}\left(  p_{\nu
}\phi\right)  +m^{2},
\end{equation}
where $p_{\mu}=\left(  E,p_{i}\right)  $. After integrating by part, we have
\begin{equation}
\mathcal{L}_{s}=\phi^{+}\left(  \mathcal{O}_{s}^{\left(  0\right)  }%
+\delta\mathcal{O}_{s}\right)  \phi,
\end{equation}
where we define%
\begin{equation}
\delta\mathcal{O}_{s}=-\frac{2}{3}\beta\left[  D_{t}^{4}-\left(  D_{x}%
^{4}+D_{y}^{4}+D_{z}^{4}\right)  \right]  +\mathcal{O}\left(  \beta
^{2}\right)  .
\end{equation}
In order to get eigenfunctions and eigenvalues of $\mathcal{O}_{s}^{\left(
0\right)  }+\delta\mathcal{O}_{s}$, we write the eigenfunctions in the form
$\phi=\exp\left(  i\mathbf{k}_{\perp}\cdot\mathbf{r}_{\perp}+ik_{z}z\right)
\sigma\left(  t\right)  $. For $\sigma\left(  t\right)  $, $\delta
\mathcal{O}_{s}$ becomes%
\begin{equation}
\delta\mathcal{O}_{s}=-\frac{2}{3}\beta\left[  \partial_{t}^{4}-k_{\perp}%
^{4}-\left(  k_{z}+eEt\right)  ^{4}\right]  +\mathcal{O}\left(  \beta
^{2}\right)  .\label{eq:DeltaOs}%
\end{equation}
Rotating to imaginary time $\tau$ and $\tilde{E}$, we can use eqn. $\left(
\ref{eq:Tao}\right)  $ and eqn. $\left(  \ref{eq:DTao}\right)  $ to write
$\delta\mathcal{O}_{s}$ in terms of ladder operators%
\begin{equation}
\delta\mathcal{O}_{s}=-\frac{1}{6}\beta e^{2}\tilde{E}^{2}\left[  \left(
a^{+}-a\right)  ^{4}-\frac{4k_{\perp}^{4}}{e^{2}\tilde{E}^{2}}-\left(
a^{+}+a\right)  ^{4}\right]  +\mathcal{O}\left(  \beta^{2}\right)  .
\end{equation}
Treating $\delta\mathcal{O}_{s}$ as perturbations, we find the first-order
correction to $\lambda_{n,k_{\perp}}^{s\left(  0\right)  }$%
\begin{equation}
\delta\lambda_{n,k_{\perp}}^{s}=\left\langle n\left\vert \delta\mathcal{O}%
_{s}\right\vert n\right\rangle =\frac{2\beta k_{\perp}^{4}}{3}%
,\label{eq:DeltaLamdaS}%
\end{equation}
where $\left\vert n\right\rangle $ is $n$th eigenstate for the one-dimensional
harmonic oscillator. The corresponding degeneracy stays same, namely
$\frac{eETVdk_{\perp}}{\left(  2\pi\right)  ^{3}}$. Therefore, eqn. $\left(
\ref{eq:EtaScalar}\right)  $ gives to $\mathcal{O}\left(  \beta\right)  $
\begin{equation}
\xi_{s}^{E}\approx-\frac{eEVT}{\left(  2\pi\right)  ^{3}}\int dk_{\perp}%
\int_{0}^{+\infty}\frac{ds}{s}\exp\left[  -\left(  k_{\perp}^{2}+m^{2}%
+\frac{2\beta k_{\perp}^{4}}{3}\right)  s\right]  \sum\limits_{n=0}^{\infty
}\exp\left[  -i\left(  2n+1\right)  eEs\right]  .
\end{equation}
Note that the above equation is the same as eqn. $\left(  \ref{eq:ESscalar}%
\right)  $ except the prefactor $\exp\left(  -\frac{2\beta k_{\perp}^{4}}%
{3}s\right)  $. Following calculations for $\gamma_{s}^{\left(  0\right)  }$,
one obtains to $\mathcal{O}\left(  \beta\right)  $
\begin{align}
\gamma_{s} &  =\operatorname{Re}\left(  \xi_{s}^{E}-\xi_{s}^{0}\right)
\nonumber\\
&  \approx\frac{eEVT}{16\pi^{3}}\int dk_{\perp}\sum_{k=1}^{\infty}\left(
-1\right)  ^{k+1}\frac{1}{k}\exp\left[  -\left(  k_{\perp}^{2}+m^{2}%
+\frac{2\beta k_{\perp}^{4}}{3}\right)  \frac{k\pi}{eE}\right]  \nonumber\\
&  \approx\frac{e^{2}E^{2}VT}{16\pi^{3}}\sum_{k=1}^{\infty}\frac{\left(
-1\right)  ^{k+1}}{k^{2}}\exp\left(  -\frac{m^{2}\pi k}{eE}-\frac{4\beta
eE}{3\pi k}\right)  .
\end{align}

\section{Fermion Pair Production}

\label{Sec:FPP}

The procedure in Section \ref{Sec:SPP} can also be applied to calculations for
a fermion field. However, two differences should be noted. First, the
Grassmann numbers are used in fermion case. Second, instead of $\mathcal{O}%
_{f}$, we usually calculate eigenvalues of $\mathcal{O}_{f}\mathcal{\tilde{O}%
}_{f}$, where $\mathcal{\tilde{O}}_{f}$ is defined as follows. In even
dimension, there exist two charge conjugate operators $C_{+}$ and $C_{-}$ such
that
\begin{equation}
C_{\pm}\gamma^{\mu}C_{\pm}^{-1}=\pm\gamma^{\mu T},
\end{equation}
where $\gamma^{\mu}$ are Gamma matrices. We then define
\begin{equation}
\mathcal{\tilde{O}}_{f}\equiv C_{-}^{-1}\left(  \gamma^{0}\left(
C_{+}\mathcal{O}_{f}C_{+}^{-1}\right)  ^{\ast}\gamma^{0}\right)  ^{T}%
C_{-}.\label{eq:OTildal}%
\end{equation}
Since $\mathcal{\tilde{O}}_{f}$ is Hermitian, one finds
\begin{equation}
\det\mathcal{\tilde{O}}_{f}\mathcal{=}\det\mathcal{O}_{f}^{\ast}%
=\det\mathcal{O}_{f}^{+}=\det\mathcal{O}_{f}.
\end{equation}
Assume the Lagrangian for a fermion field $\psi$ is give in the form of
$\mathcal{L=}\overline{\psi}\mathcal{O}_{f}\psi$. Defining eigenfunctions
$\psi_{n}$ with eigenvalues $\lambda_{n}^{\prime},$%
\begin{equation}
\mathcal{O}_{f}\psi_{n}=\lambda_{n}^{\prime}\psi_{n},
\end{equation}
we expand%
\begin{equation}
\psi=\sum\limits_{n}\xi_{n}\psi_{n}.
\end{equation}
Thus, the vacuum to vacuum amplitude for $\psi$ is
\begin{align}
e^{-\xi_{f}} &  \equiv\left\langle vac\rightarrow vac\right\rangle \propto\int
d\xi_{n}d\xi_{n}^{\ast}\exp\left(  i\int dx^{4}\lambda_{n}^{\prime}\left\vert
\xi_{n}\right\vert ^{2}\right)  \nonumber\\
&  \propto\prod\limits_{n}\lambda_{n}^{\prime}=\det\mathcal{O}_{f}%
\mathcal{=}\left(  \det\mathcal{O}_{f}\mathcal{\tilde{O}}_{f}\right)
^{\frac{1}{2}}=\exp\left[  \frac{1}{2}\text{Tr}\left(  \ln\mathcal{O}%
_{f}\mathcal{\tilde{O}}_{f}\right)  \right]  =\exp\left(  \frac{1}{2}%
\sum\limits_{n}\ln\lambda_{n}\right)
\end{align}
where $\lambda_{n}$ are eigenvalues of $\mathcal{O}_{f}\mathcal{\tilde{O}}%
_{f}$. The real part of $\xi_{f}$ is given by
\begin{equation}
\gamma_{f}=\operatorname{Re}\xi_{f}=\operatorname{Re}\left(  \xi_{f}^{E}%
-\xi_{f}^{0}\right)
\end{equation}
where $\xi_{f}^{0}$ is $\xi_{f}^{E}$ with $E=0$ and we define
\begin{equation}
\xi_{f}^{E}\equiv\frac{1}{2}\sum\limits_{n}\int_{0}^{+\infty}\frac{ds}%
{s}e^{-\lambda_{n}s}.\label{etafermion}%
\end{equation}
Squaring $\gamma_{f}$, we observe that the total pair production rate per unit
volume is simply%
\begin{equation}
\frac{\text{prob}_{\text{pair}}}{VT}=\frac{1}{VT}\left(  1-e^{-2\gamma_{f}%
}\right)  \approx\frac{2\gamma_{f}}{VT}.
\end{equation}

\subsection{Usual Fermion Field}

The Lagrangian for a charged spinor of mass $m$ and charge $e$ is
\begin{equation}
\mathcal{L}=i\overline{\psi}D^{\mu}\gamma_{\mu}\psi-m\overline{\psi}%
\psi=\overline{\psi}\mathcal{O}_{f}^{\left(  0\right)  }\psi,
\end{equation}
where $\mathcal{O}_{f}^{\left(  0\right)  }\mathcal{=}iD^{\mu}\gamma_{\mu}-m.$
Using eqn. $\left(  \ref{eq:OTildal}\right)  $, one finds%
\begin{equation}
\mathcal{O}_{f}^{\left(  0\right)  }\mathcal{\tilde{O}}_{f}^{\left(  0\right)
}=\kappa_{1}+\kappa_{2},
\end{equation}
where $\sigma^{\mu\nu}=\frac{i}{2}\left[  \gamma^{\mu},\gamma^{\nu}\right]  $,
$\left[  D_{\mu},D_{\nu}\right]  =ieF_{\mu\nu}$, $\kappa_{1}\equiv D^{2}%
+m^{2}$ and $\kappa_{2}=$ $\frac{e}{2}F_{\mu\nu}\sigma^{\mu\nu}$. Here the
eigenvalues of $\kappa_{1}$ in the case of a constant electric field were
determined in Section \ref{Sec:SPP} and are given by%
\begin{equation}
\lambda_{1}=k_{\perp}^{2}+m^{2}+\left(  2n+1\right)  e\tilde{E}.
\end{equation}
In chiral representation, we have
\begin{equation}
\kappa_{2}=ieE%
\begin{pmatrix}
-\sigma^{3} & 0\\
0 & \sigma^{3}%
\end{pmatrix}
,
\end{equation}
where $\mathbf{A}=-Et\mathbf{e}_{z}$ for a uniform electric field$\ \mathbf{E}%
=E\mathbf{e}_{z}$. Temporarily rotating to imaginary time as before, we find
that the eigenvalues of $\mathcal{O}_{f}^{\left(  0\right)  }\mathcal{\tilde
{O}}_{f}^{\left(  0\right)  }$ are
\begin{equation}
\lambda_{n,k_{\perp}}^{f\left(  0\right)  }=k_{\perp}^{2}+m^{2}+2ne\tilde
{E}\text{ for\ }n=0,1,2\cdots,
\end{equation}
and the corresponding degeneracies are%
\begin{equation}
\frac{2g_{n}eETVdk_{\perp}}{\left(  2\pi\right)  ^{3}},
\end{equation}
where $g_{0}=1$ and $g_{n>0}=2$. Thus one finds%
\begin{equation}
\xi_{f}^{E}=\frac{eEVT}{\left(  2\pi\right)  ^{3}}\int dk_{\perp}\int
_{0}^{+\infty}\frac{ds}{s}\exp\left[  -\left(  k_{\perp}^{2}+m^{2}\right)
s\right]  \sum\limits_{n=0}^{\infty}g_{n}\exp\left(  -2ineEs\right)  .
\end{equation}
Using Dirac comb, we get
\begin{equation}
\operatorname{Re}\xi_{f}^{E}=\frac{eEVT}{\left(  2\pi\right)  ^{3}}\int
dk_{\perp}\int_{0}^{+\infty}\frac{ds}{s}\exp\left[  -\left(  k_{\perp}%
^{2}+m^{2}\right)  s\right]  \pi\sum_{k=-\infty}^{\infty}\delta\left(
eEs-k\pi\right)  ,
\end{equation}%
\begin{equation}
\operatorname{Re}\xi_{f}^{0}=\frac{VT}{\left(  2\pi\right)  ^{3}}\int
dk_{\perp}\int_{0}^{+\infty}\frac{ds}{s}\exp\left[  -\left(  k_{\perp}%
^{2}+m^{2}\right)  s\right]  \pi\delta\left(  s\right)  .
\end{equation}
Therefore, we have%
\begin{align}
\gamma_{f}^{\left(  0\right)  } &  =\operatorname{Re}\left(  \xi_{f}^{E}%
-\xi_{f}^{0}\right)  \nonumber\\
&  =\frac{e^{2}E^{2}VT}{8\pi^{3}}\sum_{k=1}^{\infty}\frac{1}{k^{2}}\exp\left(
-m^{2}\frac{k\pi}{eE}\right)  .
\end{align}

\subsection{Minimal Length Modified Fermion Field}

For this case, the Lagrangian incorporating eqn. $\left(  \ref{eq:MomentumGUP}%
\right)  $ and eqn. $\left(  \ref{eq:EnergyGUP}\right)  $ for a charged spinor
$\psi$ of mass $m$ and charge $e$ can be written as
\begin{equation}
\mathcal{L}_{f}=\overline{\psi}\left(  ip_{\mu}\gamma^{\mu}-m\right)  \psi,
\end{equation}
where $p_{\mu}=\left(  E,p_{i}\right)  $. We then find%
\begin{equation}
\mathcal{L}_{f}=\overline{\psi}\mathcal{O}_{f}\psi=\overline{\psi}\left(
\mathcal{O}_{f}^{\left(  0\right)  }+\delta\mathcal{O}_{f}^{\prime}\right)
\psi,
\end{equation}
where we define
\begin{equation}
\delta\mathcal{O}_{f}^{\prime}=-\frac{i}{3}\beta D_{\mu}^{3}\gamma^{\mu
}+\mathcal{O}\left(  \beta^{2}\right)  .\label{eq:DeltaOf}%
\end{equation}
Eqn. $\left(  \ref{eq:OTildal}\right)  $ and eqn. $\left(  \ref{eq:DeltaOf}%
\right)  $ gives the product of $\mathcal{O}_{f}$ and $\mathcal{\tilde{O}}%
_{f}$ to $\mathcal{O}\left(  \beta\right)  $
\begin{equation}
\mathcal{O}_{f}\mathcal{\tilde{O}}_{f}=\mathcal{O}_{f}^{\left(  0\right)
}\mathcal{\tilde{O}}_{f}^{\left(  0\right)  }+\delta\mathcal{O}_{f}%
,\label{eq:OfOf}%
\end{equation}
where we find%
\begin{equation}
\delta\mathcal{O}_{f}=\delta\mathcal{O}_{s}-e\beta F_{\mu\nu}D_{\mu}^{2}%
\sigma^{\mu\nu}.\label{eq:DeltaOF}%
\end{equation}
The first term in eqn. $\left(  \ref{eq:DeltaOF}\right)  $ is just
$\delta\mathcal{O}_{s}$ given in eqn. $\left(  \ref{eq:DeltaOs}\right)  $,
while one can express the second term in terms of ladder operators after
rotating to imaginary time. In fact, using eqn. $\left(  \ref{eq:Tao}\right)
$ and eqn. $\left(  \ref{eq:DTao}\right)  $ gives
\begin{equation}
F_{\mu\nu}D_{\mu}^{2}\sigma^{\mu\nu}=E^{2}%
\begin{pmatrix}
-\sigma^{3} & 0\\
0 & \sigma^{3}%
\end{pmatrix}
\left(  a^{+2}+a^{2}\right)  ,
\end{equation}
which doesn't contribute to the first-order correction to eigenvalues
$\lambda_{n,k_{\perp}}^{f\left(  0\right)  }$ of the leading operator
$\mathcal{O}_{f}^{\left(  0\right)  }\mathcal{\tilde{O}}_{f}^{\left(
0\right)  }$ in eqn. $\left(  \ref{eq:OfOf}\right)  $ since $\left\langle
n\left\vert F_{\mu\nu}D_{\mu}^{2}\sigma^{\mu\nu}\right\vert n\right\rangle $
is zero. Thus, the eigenvalues of $\mathcal{O}_{f}\mathcal{\tilde{O}}_{f}$ to
$\mathcal{O}\left(  \beta\right)  $ are%
\begin{equation}
\lambda_{n,k_{\perp}}^{f}\approx k_{\perp}^{2}+m^{2}+2ne\tilde{E}\text{
}+\frac{2\beta k_{\perp}^{4}}{3}\text{ for\ }n=0,1,2\cdots,
\end{equation}
where we use eqn. $\left(  \ref{eq:DeltaLamdaS}\right)  $. The corresponding
degeneracies are also%
\begin{equation}
\frac{2g_{n}eETVdk_{\perp}}{\left(  2\pi\right)  ^{3}},
\end{equation}
where $g_{0}=1$ and $g_{n>0}=2$. Therefore, one finds for $\xi_{f}^{E}$%
\begin{equation}
\xi_{f}^{E}\approx\frac{eEVT}{\left(  2\pi\right)  ^{3}}\int dk_{\perp}%
\int_{0}^{+\infty}\frac{ds}{s}\exp\left[  -\left(  k_{\perp}^{2}+m^{2}%
+\frac{2\beta k_{\perp}^{4}}{3}\right)  s\right]  \sum\limits_{n=0}^{\infty
}g_{n}\exp\left(  -2ne\tilde{E}s\right)  .
\end{equation}
Using Dirac Comb, we find that
\begin{equation}
\operatorname{Re}\xi_{f}^{E}\approx\frac{eEVT}{\left(  2\pi\right)  ^{3}}\int
dk_{\perp}\int_{0}^{+\infty}\frac{ds}{s}\exp\left[  -\left(  k_{\perp}%
^{2}+m^{2}+\frac{2\beta k_{\perp}^{4}}{3}\right)  s\right]  \pi\sum
_{k=-\infty}^{\infty}\delta\left(  eEs-k\pi\right)  ,
\end{equation}%
\begin{equation}
\operatorname{Re}\xi_{f}^{0}\approx\frac{VT}{\left(  2\pi\right)  ^{3}}\int
dk_{\perp}\int_{0}^{+\infty}\frac{ds}{s}\exp\left[  -\left(  k_{\perp}%
^{2}+m^{2}+\frac{2\beta k_{\perp}^{4}}{3}\right)  s\right]  \pi\delta\left(
s\right)  .
\end{equation}
Again, we obtain $\gamma_{f}$ to $\mathcal{O}\left(  \beta\right)  $
\begin{align}
\gamma_{f} &  =\operatorname{Re}\left(  \xi_{f}^{E}-\xi_{f}^{0}\right)
\nonumber\\
&  \approx\frac{eEVT}{8\pi^{3}}\sum_{k=1}^{\infty}\frac{1}{k}\int dk_{\perp
}\exp\left[  -\left(  k_{\perp}^{2}+m^{2}+\frac{2\beta k_{\perp}^{4}}%
{3}\right)  \frac{k\pi}{eE}\right]  \nonumber\\
&  \approx\frac{e^{2}E^{2}VT}{8\pi^{3}}\sum_{k=1}^{\infty}\frac{1}{k^{2}}%
\exp\left(  -m^{2}\frac{k\pi}{eE}-\frac{4\beta eE}{3\pi k}\right)  .
\end{align}

\section{ Discussion and Conclusion}

\label{Sec:Con}

First, it is noted that the problem of scalar particles pair creation by an
electric field in the presence of a minimal length is also studied in ref.
\cite{CON-Haouat:2013yba}. The authors considered another GUP of form%
\begin{align}
x_{i} &  =x_{0i,}\\
p_{i} &  =p_{0i}\left(  1+\beta p^{2}\right)  ,
\end{align}
where $x_{0i,}$ and $p_{0i}$ satisfy the canonical commutation relations.
Using Bogoliubov transformations, they found
\begin{equation}
\gamma_{s}\sim\exp\left[  -m^{2}\frac{\pi}{eE}\left(  1+\frac{\beta m^{2}}%
{4}\left(  1-\frac{e^{2}E^{2}}{m^{4}}\right)  \right)  \right]  ,
\end{equation}
where their minimal length corrections depend on the mass of scalar particles
while our results don't, at least to $\mathcal{O}\left(  \beta\right)  $.

In the case of electron--positron pairs, the pair production is irrelevant for
laboratory electric fields, let alone the minimal length corrections. However,
the Schwinger mechanism has to do with the Unruh effect which predicts that an
accelerating observer will observe a thermal spectrum of photons and
particle--antiparticle pairs at temperature $T=\frac{a}{2\pi},$ where $a$ is
the acceleration\cite{CON-Unruh:1976db}. We now investigate how our results
imply for the corrections to the Unruh temperature due to the minimal length.
Considering a free particle of charge $e$ and mass $m$ moving in a static
electric field $E$, one finds the particle have acceleration $a=\frac{em}{E}$.
Keeping only the leading term, the pair production probability per unit volume
per unit time is%
\begin{equation}
\frac{\text{prob}_{\text{pair}}}{VT}\sim\exp\left(  -m^{2}\frac{\pi}{eE}%
-\frac{4\beta eE}{3\pi}\right)  .
\end{equation}
Identifying the reduced mass $\frac{m}{2}=K$ as the energy associated with the
pair production process, we find that the production probability can be
written in the form%
\begin{equation}
\text{prob}_{\text{pair}}\sim\exp\left[  -\frac{K}{a/2\pi}\left(
1+\frac{4\beta a^{2}}{3\pi^{2}}\right)  \right]  .\label{eq:UnruhTemp}%
\end{equation}
The minimal length modified Unruh temperature can be read from eqn. $\left(
\ref{eq:UnruhTemp}\right)  ,$ which gives%
\begin{equation}
T_{u}\sim\frac{a}{2\pi}\left(  1-\frac{4\beta a^{2}}{3\pi^{2}}\right)
,\label{eq:MUnruhTemp}%
\end{equation}
where $a$ is the acceleration of the observer.

The Unruh effects can be relevant to the phenomenon of black hole decay due to
pair creation, the Hawking radiation\cite{CON-Hawking:1974sw}. Consider a
Schwarzschild black hole with the black hole's mass, $M$. The event horizon of
the Schwarzschild black hole is $r_{h}=2M$. Noting that the gravitational
acceleration at the event horizon is given by%
\begin{equation}
a=\frac{M}{r_{h}^{2}}=\frac{1}{4M},
\end{equation}
one finds from eqn. $\left(  \ref{eq:MUnruhTemp}\right)  $ that the minimal
length modified Hawking temperature is%
\begin{equation}
T_{h}\sim\frac{1}{8\pi M}\left(  1-\frac{\beta}{12\pi^{2}M^{2}}\right)
.\label{eq:HawkingTemp}%
\end{equation}
Using the first law of the black hole thermodynamics, we find the corrected
black hole entropy is%
\begin{align}
S &  =\int\frac{dM}{T}\nonumber\\
&  \sim\frac{A}{4}+\frac{\beta}{3\pi}\ln\left(  \frac{A}{16\pi}\right)
,\label{eq:entropy}%
\end{align}
where $A=4\pi r_{h}^{2}=16\pi M^{2}$ is the area of the horizon. The
logarithmic term in eqn. $\left(  \ref{eq:entropy}\right)  $ is the well known
correction from quantum gravity to the classical Bekenstein-Hawking entropy,
which have appeared in different studies of GUP modified thermodynamics of
black holes\cite{Bina:2010ir,Chen:2002tu,Xiang:2009yq}. A careful reader might
not be satisfied with the above heuristic handwaving argument relating the
Schwinger mechanism to the Unruh temperature and the Hawking temperature.
However, using the minimal length modified Hamilton-Jacobi Method
incorporating eqn. $\left(  \ref{eq:MomentumGUP}\right)  $ and eqn. $\left(
\ref{eq:EnergyGUP}\right)  $, we found in ref. \cite{CON-Benrong:2014woa} that
the corrected Hawking temperature for the Schwarzschild black is given by
\begin{equation}
T_{h}\sim\frac{1}{8\pi M}\left(  1-\frac{\beta}{6M^{2}}\right)
,\label{eq: Temp}%
\end{equation}
which is almost same as eqn. $\left(  \ref{eq:HawkingTemp}\right)  $ except
the numerical factor in front of $\beta$. 

In this paper, incorporating effects of the minimal length, we derived the
deformed Schwinger mechanism for both scalars and fermions in a static uniform
electric field. Using handwaving argument, the implications of our results for
the Unruh temperature and the Hawking temperature are also discussed.

\noindent\textbf{Acknowledgements. }

We would like to acknowledge useful discussions with Y. He, Z. Sun and H. W.
Wu. This work is supported in part by NSFC (Grant No. 11005016, 11175039 and
11375121) and SYSTF (Grant No. 2012JQ0039).

\end{document}